\documentclass[aps, superscriptaddress,pra,twocolumn,footinbib]{revtex4-1}
\usepackage[utf8]{inputenc}
\usepackage{amsmath}
\usepackage{bm}
\usepackage{amsfonts}
\usepackage{mathtools}
\usepackage{graphicx}
\usepackage{cleveref}
\usepackage{autonum}
\usepackage{fixme}
\usepackage{color}

\renewcommand{\vec}[1]{\boldsymbol{#1}}
\renewcommand{\k}[0]{\mathbf{k}}

\newcommand{\g}[0]{\gamma}

\renewcommand{\eqref}[1]{(\ref{#1})}
\def\nn{\nonumber}

\begin{document}

\title{Constraints on the energy spectrum of non-Hermitian models in open environments}

\author{Jonatan Melk\ae r \surname{Midtgaard}}
\email{midtgaard@phys.au.dk}
\affiliation{Department of Physics and Astronomy, Aarhus University, DK-8000 Aarhus C,  Denmark}
\author{Zhigang \surname{Wu}}
\email{wuzg@sustech.edu.cn}
\affiliation{Shenzhen Institute for Quantum Science and Engineering and Department of Physics, Sustech, Shenzhen 518055, China}
\affiliation{Center for Quantum Computing, Peng Cheng Laboratory, Shenzhen 518055, China}
\author{\surname{Yu} Chen}
\email{spaceexplorer@163.com}
\affiliation{Center for Theoretical Physics and Department of Physics, Capital Normal University, Beijing, 100048, China}
\begin{abstract}
Motivated by recent progress on non-Hermitian topological band theories, we study the energy spectrum of a generic two-band non-Hermitian Hamiltonian. We prove rigorously that the complex energy spectrum of such a non-Hermitian Hamiltonian is restricted to the lower complex plane, provided that the parameters of the Hamiltonian satisfy a certain constraint. Furthermore, we consider one specific scenario where such a non-Hermitian Hamiltonian can arise, namely a two-band model coupled to an environment, and show that this aforementioned constraint orignates from very general physical considerations. Our findings are relevant in the definition of the energy gap in non-Hermitian topological band theories and also have implications on simulations of such theories using quantum systems.
\end{abstract}
\maketitle

\section{Introduction}
Topological band theory was originally proposed to describe a new type of phase of matter in electronic systems\cite{Kane10,Zhang11,Das16,Ryu16,Vishwanath18}, namely the symmetry protected topological phases. Now it has found wide applications also in photonic materials, cold atoms, as well as mechanical systems\cite{Zoller16,Spielman18,rev_photon,rev_mech}. Very recently, the topological band theory was extended to describing open systems governed by non-Hermitian Hamiltonians, where certain topological properties are found to be robust against non-Hermicity\cite{Levitov09,Levitov10,Hughes11,Kohmoto11,Kohmoto12,Nazarov13,Huang13,Sch13,Sch15,Aguado16,Molina16,Molina17,Molina18,SSHa,Lee16a,SSHb,Duan17,Menke17,Haas17,Lieu18,Zyuzin18,Fan18,Torres18,Shen,ShenQO,SSHc,ChenS18,FuArc17,Ueda18a,Khanikaev18,Fu18,SSHd,Wang,Ueda18c,SSHe,Bergholtz18a,Ueda,Notomi18,Ueda18e,Yoshida18,Yi18c,Hu18,Zhang18,Thomale18,Song18,Bergholtz18b,Yokoyama18,Ueda18g,ZhenB18c,Bergholtz18c,Bergholtz18d,Ueda18f,CZ18,Lee19,Regnault19,Nori19a,Silverinha19,WangR19,Quasicrystal,CShu19,Gilbert19,doubleGreen19,Schnyder19,Slager19,Das19,Kawabata19,Murakami19,Ge19a,Zhangcw19a,Yi19th,Fan19,Bergholtz19a,Rau19,ZhaiY19,Peled19,Zhangcw19b,Sato19,Lieu19,Kawagushi19,Gefen19,Wang19a,Wang19b,Schomerus15ex,Szameit15ex,ZhenB15ex,Szameit17ex,Obuse17ex,Yi17ex,Amo17,Fu18ex,Feng18exa,Feng18exb,Khajavikhan18ex,TopLT,TopLE,Xue18ex,Rech18ex,ZhenB19b}. A non-Hermitian version of bulk-edge correspondence can be established\cite{Shen,SSHd,Wang} and the physical connection between Hall conductance and a non-hermitian topological invariant is found\cite{CZ18}. Many new phenomena unique to the non-Hermitian systems, such as the existence of topological bulk Fermi arcs\cite{FuArc17,Fu18,Fu18ex}, non-Hermitian skin effects\cite{SSHa,SSHb,SSHc,SSHd,Wang,SSHe,Ueda,Thomale18,CShu19,ZhaiY19,Wang19a,Wang19b} and non-Hermitian knotted Fermi ribbons\cite{Bergholtz18a,Bergholtz18c} have also been discovered.

Of central importance in the non-Hermitian topological band theory is defining an energy gap, which is not as straightforward as for the Hermitian courterpart due to the fact that the energy spectrum here is generally complex. In previous studies, concepts such as point gap and line gap are introduced\cite{Ueda18e,Ueda18f,Kawabata19}; the former means that the complex spectrum does not enclose the origin of the complex energy plane while the latter depicts a scenario where the complex spectrum can be separated into two parts by a line. Classification theories of non-Hermitian bands are proposed based on whether these mathematical gaps are protected by certain symmetries\cite{Ueda18e,Ueda18f,Kawabata19}. An implicit assumption made in these discussions of the non-Hermitian energy gap, however, is that the energy spectrum of the non-Hermitian models exist in both the upper and lower complex plane.  But it is not obvious that this is always the case for any non-Hermitian system. In other words, are there any constraints on the non-Hermitian models when the physical origins of such models are taken into account? If so, what are the consequences of such constraints in relation to the energy spectrum of the non-Hermitian models?



In this paper, we address these important questions in the context of a generic two-band system coupled to an environment. We demonstrate that such a coupling generally leads to the following effective non-Hermitian Hamiltonian for the system
\begin{align}
 \mathcal{H}_{\rm eff} (\k) = \left[\boldsymbol{d}(\k) + i\boldsymbol{\gamma}(\k)\right]\cdot\vec{\sigma} + i\gamma_0(\k)\sigma_0,
 \label{calHeff}
\end{align}
where $\vec{d}(\k)$ and $\vec{\g}(\k)$ are two three-component vectors, $\vec{\sigma} = (\sigma_x,\sigma_y,\sigma_z)$ are the Pauli matrices and $\sigma_0 = \mathbb{I}$ is the identity matrix.
In deriving the effective Hamiltonian we find the following mathematical constraint for the parameters of the Hamiltonian
\begin{align}
-\gamma_0(\k) \geq  \sqrt{\g_x^2(\k)+\g_y^2(\k)+\g_z^2(\k)}.
\label{eq:condition}
\end{align}
Further, we prove rigorously that this constraint leads to a crucial property of the non-Hermitian Hamiltonian, namely that the imaginary part of all its eigenvalues must be non-positive. We will also show that these seemingly mathematical statements in fact result from very general physical considerations. Finally, as a concrete example, we consider a one-dimensional system coupled to a chain of tight-binding baths and determine explicitly the parameters in the effective Hamiltonian as well as the real space behaviour of the non-Hermitian terms.

The structure of the paper is as follows. In Section \ref{sec:tbmodel} we will begin with the non-Hermitian Hamiltonian in (\ref{calHeff}) and prove that all its eigenvalues have non-positive imaginary parts if \eqref{eq:condition} is obeyed. Next, in Section \ref{sec:secoupling} we will provide the derivation of such a non-Hermitian Hamiltonian from system-environment coupling and in doing so show that the constraint \eqref{eq:condition} is indeed satisfied. We then study the behaviour of non-hermitian terms in real space in Section \ref{sec:realspace} and provide concluding remarks in Section \ref{sec:conclusion}.

\section{Properties of the eigenvalues for a two-band non-Hermitian model}
\label{sec:tbmodel}

We consider the following two-band non-hermitian effective hamiltonian
\begin{align}
\hat H_{\rm eff} &= \sum_{\k}  \vec{\psi}^\dagger_{\k} \mathcal{H}_{\rm eff}(\k)\vec{\psi}_{\k}
\label{eq:fullham}
\end{align}
where $\mathcal{H}_{\rm eff}(\k)$ is given by (\ref{calHeff}), the summation is over the crystal momenta in the first Brillouin zone and $\vec{\psi}_\k = (\psi_{\uparrow,\k},\,\psi_{\downarrow,\k})^{\rm T}$  is a two-component spinor.

Such an effective Hamiltonian has been used in most of the studies on topological phases of non-Hermitian systems. We will prove an important property of this Hamiltonian, namely that all its complex eigenvalues must have a non-positive imaginary part, provided that the inequality (\ref{eq:condition})
is satisfied for all $\k$.  In the next section, we will show that the constraint (\ref{eq:condition}) in fact follows naturally from general physical considerations.

We denote the eigenvalues of the non-Hermitian matrix $\mathcal{H}_{\rm eff} (\k)$ in (\ref{eq:fullham}) by $\lambda^{\pm}(\k)$, which are determined by
\begin{align}
{\rm det} \left ( \lambda^{\pm}(\k)\mathbb{I} -  \mathcal{H}_{\rm eff} (\k) \right) = 0.
\end{align}
 The imaginary part of these eigenvalues can be written as
 \begin{align}
 \text{Im}\lambda^{\pm}(\k) = \gamma_0 (\k)\pm M(\k)
 \label{Imlambda}
 \end{align}
  where
\begin{align}
M = \left[(|\vec{d}|^2-|\vec{\gamma}|^2)^2 + (2\vec{d}\cdot\vec{\gamma})^2 \right]^{1/4} \sin\varphi
\label{Mdef}
\end{align}
with
\begin{align}
\varphi =\frac{1} {2}\tan^{-1}\left(\frac{2\vec{d}\cdot\vec{\gamma}}{|\vec{d}|^2-|\vec{\gamma}|^2}\right ).
\label{tantheta}
\end{align}
Defining the function $f = \left[ \left(|\vec{d}|^2-|\vec{\gamma}|^2\right)^2 + (2\vec{d}\cdot\vec{\gamma})^2 \right]^{1/4}$, we find from (\ref{tantheta})
\begin{align}
\sin^2\varphi
= \frac{1}{2} \left( 1- \frac{|\vec{d}|^2 - |\vec{\gamma}|^2}{f^2} \right).
\end{align}
Substituting this expression in (\ref{Mdef}) we obtain
\begin{align}
{|M|} = {{f}|\sin\varphi|}= \frac{1}{\sqrt{2}} \sqrt{ {f^2} - {|\vec{d}|^2} + |\vec{\gamma}|^2  }.
\end{align}
Since
\begin{equation}
{f^2}\leq  \left[ (|\vec{d}|^2-|\vec{\gamma}|^2)^2 + 4|\vec{d}|^2 |\vec{\gamma}|^2 \right]^{1/2} = |\vec{d}|^2 +|\vec{\gamma}|^2,
\label{eq:FoG2}
\end{equation}
we have
\begin{align}
{|M(\k)|} \leq |\vec{\gamma} (\k)|.
\label{Mleqg}
\end{align}
Together with the condition (\ref{eq:condition}), (\ref{Mleqg}) implies that
\begin{align}
 \text{Im}\lambda^{\pm}(\k)  \leq 0,
 \label{eq:property}
\end{align}
namely all the eigenvalues of $\mathcal{H}_{\rm eff} (\k)$ have a non-positive imaginary part. In other words, the complex eigenvalues of the non-Hermitian Hamiltonian must all lie in the lower half complex plane. We emphasise that the above analysis is a rigorous mathematical discussion on the property of a generic two-band non-Hermitian Hamiltonian and the conclusion stands regardless of the origin of this Hamiltonian. In the following section, we will discuss the origin of (\ref{eq:condition}) and the physical implication of (\ref{eq:property}) in the context of a system coupled to an environment.

\section{Construction of the non-Hermitian model from System-Environment Coupling}
\label{sec:secoupling}
In this section, we show how the effective non-Hermitian Hamiltonian (\ref{eq:fullham}) can be obtained by considering a two-band system coupled to an environment. By tracing out the degrees of freedom of the environment within certain approximations, we not only derive the effective Hamiltonian (\ref{eq:fullham})  but also prove that the constraint in (\ref{eq:condition}) is always fulfilled. To be specific, we consider a general two-band model described by the Hamiltonian
\begin{align}
\hat{H}_{\rm sys}=\sum_\k \vec{\psi}_\k^\dag \left [ {\bf d}_0(\k)\cdot\vec{\sigma} \right ]\vec{\psi}_\k = \sum_\k \vec{\psi}_\k^\dag \mathcal{H}_\text{sys} (\k)\vec{\psi}_\k,
\end{align}
where ${\bf d}_0(\k)$ is a three-component vector. Such a model is realised in a bipartite lattice with $N$ unit cells and the up and down arrows in  $\vec{\psi}_\k = (\psi_{\uparrow,\k},\,\psi_{\downarrow,\k})^{\rm T}$ are used to distinguish the sites within a unit cell. The particles on each site are locally coupled to an individual bath with many internal modes. In addition, the particles in the same internal mode can hop between the baths such that the collection of the baths forms the environment described by the Hamiltonian
\begin{equation}
\hat H_\text{bath} = \sum_{\k,\alpha} \vec{b}^\dagger_{\k,\alpha} \mathcal{H}_{\rm bath} \vec{b}_{\k,\alpha}
\label{Hbath}
\end{equation}
where $\vec{b}_{\k,\alpha} = (b_{\uparrow,\k,\alpha},\, b_{\downarrow,\k,\alpha})^{\rm T}$  and
\begin{align}
\mathcal{H}_{\rm bath} = \begin{pmatrix}
\omega_\alpha & J_\k \\
J_\k^* & \omega_\alpha
\end{pmatrix}.
\label{Hb}
\end{align}
Here $\omega_\alpha$ is the mode energy in the bath and
\begin{align}
J_\k = |J_\k| e^{i\phi_\k}
\end{align}
 describes the coupling between the baths. $\mathcal{H}_{\rm bath} $ can be diagonalised by an unitary transformation $U_{\k} $
 \begin{align}
\label{Ukdef}
    U_\k &= \frac{\sqrt{2}}{2}
    \begin{pmatrix}
    e^{i\phi_\k} & -e^{i\phi_\k} \\
    1 & 1
    \end{pmatrix},
\end{align}
such that
\begin{align}
\hat{H}_{\rm bath}=\sum_{\k,\alpha}\vec \beta_{\bf k,\alpha}^\dag
\begin{pmatrix}
\epsilon_{\k,\alpha}^+ & 0 \\
0 & \epsilon_{\k,\alpha}^-
\end{pmatrix}
\vec \beta_{\bf k,\alpha},
\end{align}
where $\vec \beta_{\bf k,\alpha} = U^\dag_{\k} \vec{b}_{\k,\alpha}$ and
\begin{align}
\epsilon_{\k,\alpha}^{\pm} = \omega_\alpha \pm |J_\k|.
\end{align}
 Finally, since the system-bath coupling is local, the part of the Hamiltonian describing the interaction between the system and the environment can be written as
\begin{align}
\hat{H}_{\rm int}=\sum_{ \bf n,\alpha}\left [ \vec \psi_{\bf n}^\dag K \vec b_{ \bf n,\alpha}^{} + \text{h.c.} \right ],
\end{align}
where $ \bf n$ labels the unit cells, $K = {\rm diag} (\kappa_1, \kappa_2)$ contains real coupling strengths $\kappa_1$ and $\kappa_2$, and
\begin{align}
\vec b_{\bf n,\alpha} = \frac{1}{\sqrt{N}}\sum_\k e^{i\k\cdot \bf R_{\bf n}}\vec b_{\k,\alpha}  \\
\vec \psi_{\bf n} = \frac{1}{\sqrt{N}}\sum_\k e^{i\k\cdot \bf R_{\bf n}}\vec \psi_{\k} .
\end{align}
Here $\bf R_{\bf n}$ denotes the lattice vector. Transforming back to Fourier space and adopting the diagonalised basis for the baths, we obtain
\begin{align}
\hat{H}_{\rm int}&=\sum_{\bf k,\alpha}\left [\vec{\psi}^\dag_{\k}K \vec b_{\bf k,\alpha}^{} + \text{h.c.}\right ]  \nn \\
&= \sum_{\bf k,\alpha}\left [\vec{\psi}^\dag_{\k}K U^\dag_{\k}\vec\beta_{\k,\alpha}^{} + \text{h.c.}\right ].
\end{align}
The full Hamiltonian for the system-environment is thus given by
\begin{equation}
\hat{H}_{\rm f}=\hat{H}_{\rm sys}+\hat{H}_{\rm bath}+\hat{H}_{\rm int}.
\label{eq:hamthreeterms}
\end{equation}

To obtain an effective Hamiltonian from \eqref{eq:hamthreeterms}, we study the retarded Green's function for the system defined as
\begin{align}
&G^{\rm R}(\k,t-t') = -i\Theta(t-t')  \nn \\
\times&\begin{pmatrix}
\left\langle\left \{\psi_{\uparrow,\k}(t),\psi_{\uparrow,\k}^\dag(t') \right \} \right \rangle  &\left \langle\left \{\psi_{\uparrow,\k}(t),\psi_{\downarrow,\k}^\dag(t') \right \} \right \rangle  \\
\left \langle\left \{\psi_{\downarrow,\k}(t),\psi_{\uparrow,\k}^\dag(t') \right \} \right \rangle & \left \langle\left \{\psi_{\downarrow,\k}(t),\psi_{\downarrow,\k}^\dag(t') \right \} \right \rangle
\end{pmatrix},
\end{align}
where $\psi_{\uparrow,\k}(t) \equiv e^{i\hat H_{\rm f} t} \psi_{\uparrow,\k}e^{-i\hat H_{\rm f} t}$ and $\langle \cdots \rangle$ stands for the average over the system and the environment.
Using the standard Keldysh formalism, we integrate out the degrees of freedom of the baths and obtain the Dyson's equation for the retarded Green's function in the momentum-frequency space as
\begin{align}
[{G}^\text{R}(\k,\epsilon)]^{-1}&=[{G}^\text{R}_0(\k,\epsilon)]^{-1}- \Sigma^{\rm R}(\k,\epsilon).
\label{GR}
\end{align}
Here ${G}^\text{R}_0(\k,\epsilon)$  is the Green's function in the absence of coupling to the baths, whose inverse is given by
\begin{align}
[{G}^\text{R}_0(\k,\epsilon)]^{-1} = \epsilon\mathbb{I} -{\cal H}_\text{sys}(\k).
\end{align}
 $\Sigma^{\rm R}(\k,\epsilon)$ is the self-energy given by
\begin{align}
\Sigma^{\rm R}(\k,\epsilon) = KU_{\bf k}\left[ \sum_\alpha\Pi_{0,\alpha}^\text{R} (\k,\epsilon) \right ]U^\dag_{\bf k}K^\dag,
\label{Sigmarf}
\end{align}
where $\Pi_{0,\alpha}^\text{R}(\k,\epsilon)=[\epsilon\mathbb{I} -\rm{diag}(\epsilon_{\k,\alpha}^+,\epsilon_{\k,\alpha}^-)+i0^+]^{-1}$ is the retarded Green's function for the bath  mode $\alpha$.
Introducing the spectral function $A({\k},\epsilon)={\rm diag}(A_+({\k},\epsilon),A_-({\k},\epsilon))$, where
\begin{equation}
A_{\pm}({\k},\epsilon)\equiv 2\pi \sum_{\alpha}\delta(\epsilon-\epsilon_{\k,\alpha}^\pm),
\label{Apmdef}
\end{equation}
we can write
\begin{align}
\sum_\alpha\Pi^{\rm R}_{0,\alpha}(\k,\epsilon)&=\int \frac{d\omega}{2\pi}\frac{A({\bf k},\omega)}{\epsilon-\omega+i0^+} \nn \\
& = \mathcal {P}\int \frac{d\omega}{2\pi}\frac{A({\bf k},\omega)}{\epsilon-\omega} - i\frac 1 2 A({\bf k},\epsilon).
\end{align}
Thus $\Sigma^{\rm R}(\k,\epsilon)$ contains a dispersion (real) part and a dissipation (imaginary) part. We shall assume that  the self-energy is slowly varying in $\epsilon$ around the Fermi energy $\mu$, whereby we can approximate it to be independent of $\epsilon$, i.e.,
\begin{align}
\Sigma^{\rm R}(\k,\epsilon)  \approx \Sigma^{\rm R}(\k,\mu).
\label{sigmaapprox}
\end{align}
This is a good approximation for weak couplings between the environment and the system. Under this approximation we can write
\begin{equation}
\Sigma^{\rm R} (\k,\epsilon) \approx \mathcal {H}_{\rm disp}  + i \mathcal{H}_{\rm diss},
\end{equation}
where
\begin{align}
\label{Hdisp}
\mathcal {H}_{\rm disp} & = KU_{\bf k}^\dag\left[ \mathcal {P}\int \frac{d\omega}{2\pi}\frac{A({\bf k},\omega)}{\mu-\omega}  \right ]U_{\bf k}K^\dag \\
 \mathcal{H}_{\rm diss} & = -\frac{1}{2}KU_{\bf k} A({\bf k},\mu)U^\dag_{\bf k}K^\dag.
 \label{Hdiss}
\end{align}

Now in order to extract an effective Hamiltonian for the system, we recall that in the absence of coupling to the baths, the single particle Hamiltonian of the system can be obtained directly from (\ref{GR}) as
\begin{align}
\mathcal H_\text{sys} =\epsilon\mathbb{I} - [{G}^\text{R}_0(\k,\epsilon)]^{-1}.
\end{align}
Analogously, in the presence of the baths we obtain the effective single particle Hamiltonian as
\begin{align}
\mathcal H_{\rm eff}& = \epsilon\mathbb{I} - [{G}^\text{R}(\k,\epsilon)]^{-1} \nn \\
& \approx \mathcal {H}_\text{sys} + \mathcal {H}_{\rm disp}  + i \mathcal{H}_{\rm diss}.
\label{Heffdef}
\end{align}
By writing
\begin{align}
 \mathcal {H}_\text{sys} + \mathcal {H}_{\rm disp} = {\bf d}(\k)\cdot\vec{\sigma}
\end{align}
and
\begin{equation}
\mathcal{H}_\text{diss} =\gamma_0 \sigma_0 + \sum_{i=x,y,z} \gamma_i \sigma_i,
\label{Hdecomp}
\end{equation}
we arrive at the effective Hamiltonian given in (\ref{eq:fullham}).

We next show that the previously assumed constraint (\ref{eq:condition}) follows from the physical consideration that the baths have finite density of states. From (\ref{Hdecomp}) we can calculate the $\gamma_i$ coefficients as
\begin{align}
    \gamma_i &= \frac{1}{2} \text{Tr}(\sigma_i \mathcal{H}_\text{diss})
    \label{gamma_def}
\end{align}
where $i=0,x,y,z$. Using (\ref{Hdiss}) and (\ref{Ukdef}) in (\ref{gamma_def}) we find
\begin{align}
\label{gamma0}
\gamma_0(\k) &= -\frac{\kappa_1^2+\kappa_2^2}{8}\left [ A_{+}(\k,\mu)+A_{-}(\k,\mu) \right ]\\
\label{gammax}
\gamma_x(\k) &= -\frac{\kappa_1\kappa_2 \cos {\phi_\k}}{4}\left [A_{+}(\k,\mu) - A_{-}(\k,\mu) \right ]\\
\label{gammay}
\gamma_y(\k) &= -\frac{\kappa_1\kappa_2 \sin {\phi_\k}}{4}\left [ A_{+}(\k,\mu) - A_{-}(\k,\mu)\right  ]\\
\label{gammaz}
\gamma_z(\k) &= - \frac{\kappa_1^2-\kappa_2^2}{8}\left  [A_{+}(\k,\mu)+A_{-}(\k,\mu) \right ].
\end{align}

It is clear from (\ref{gamma0}) that $\gamma_0$ is always non-positive because the spectral functions of the baths are non-negative. A straightforward calculation yields
\begin{align}
    \frac{\gamma_x^2 + \gamma_y^2 + \gamma_z^2}{\gamma_0^2} = 1-
   \frac{16 {\kappa_2}^2{\kappa_1}^2 A_+ A_-}{(\kappa_1^2+\kappa_2^2)^2(A_+  +  A_-)^2}.
    \label{eq:gammas}
\end{align}
Again, since the spectral functions of the bath are non-negative, it is easy to show from the above expression that
\begin{align}
0 \leq  \frac{\gamma_x^2 + \gamma_y^2 + \gamma_z^2}{\gamma_0^2} \leq 1
\end{align}
This result, along with the fact that $\gamma_0$ is non-positive, immediately leads to the constraint in (\ref{eq:condition}).

In the previous section we have proven that the constraint in (\ref{eq:condition}) means that the imaginary part of the eigenvalues of the effective non-Hermitian Hamiltonian must be non-positive. Now we are in a position to discuss the physical implication of this property.  From (\ref{Heffdef}) we can write
\begin{align}
{G}^\text{R}(\k,\epsilon) = \big (\epsilon\mathbb{I} - \mathcal {H}_{\rm eff}(\k) +i 0^+ \big)^{-1}.
\end{align}
The density of states of the system is thus given by
\begin{align}
D_{\rm eff}(\epsilon) &= -\frac{1}{\pi N\Omega}\sum_\k {\rm Tr} \left [ {\rm Im} {G}^\text{R}(\k,\epsilon)\right ] \nonumber \\
& = -\frac{1}{\pi N\Omega} \sum_{\k}\sum_{s=\pm} \frac{\text{Im}\lambda^s (\k)}{(\epsilon-\text{Re}\lambda^{s} (\k))^2 + (\text{Im}\lambda^{s}(\k))^2}.
\end{align}
We now see that the property in (\ref{eq:property}) ensures that the density of states of the system is non-negative, which is a physical requirement for any non-Hermitian effective Hamiltonian.

\section{Real space behaviour of the Non-Hermitian Terms}
\label{sec:realspace}
\begin{figure}[htb]
\centering
\includegraphics[width=\columnwidth]{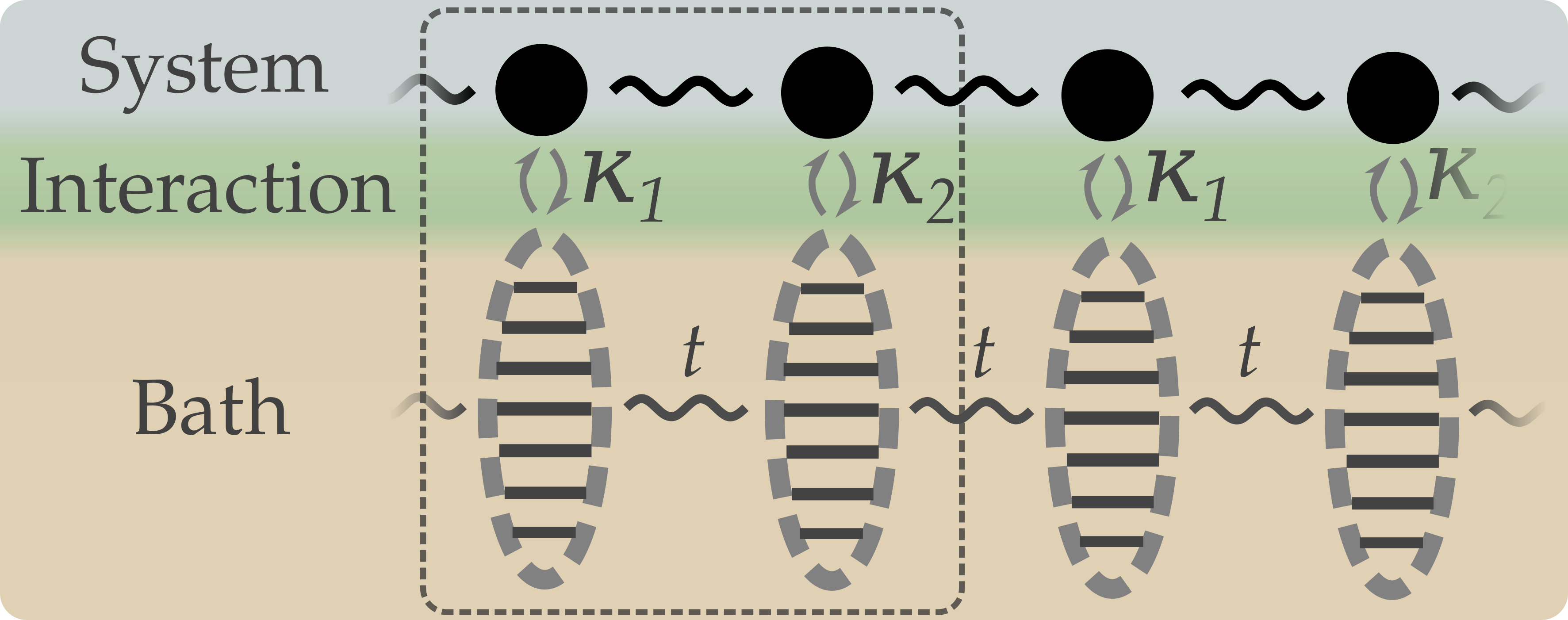}
\caption{(Color online) One-dimensional example of a SSH chain, coupled to an environment composed of tight-binding local baths. The unit cell is comprised of two sites, yielding a two-band model. }
\label{fig:system}
\end{figure}
In the previous section we have seen that the $\gamma_i$ parameters in the non-Hermitian effective Hamiltonian are momentum dependent, indicating that the dissipation of the system is generally non-local in real space. Although this is to be expected because the baths are coupled to each other, it is instructive to
examine a specific example and study how the properties of the baths affect the real space behaviour of the non-hermitian terms. In this section, we consider a two-band chain (e.g. a Su-Schrieffer-Heeger (SSH) model) with a bath at every lattice site, as illustrated in Fig. \ref{fig:system}. The coupling between the baths is described by (\ref{Hb}). For simplicity, we take the tight-binding coupling between the baths for which
\begin{align}
J_k = -t(1+e^{ika}),
\end{align}
 where $t$ is the nearest neighbour hopping parameter, $a$ is the spacing between the adjacent unit cells and $-\pi/a<k\leq \pi/a$ is the wave vector in the first Brillouin zone. We thus find the bath band dispersion is  given by
 \begin{align}
 \epsilon_{k,\alpha}^{\pm} = \omega_\alpha \pm 2t \cos \frac {ka}{2}.
 \end{align}
 and the phase of the coupling is determined by
 \begin{align}
e^{i \phi_k} =e^{i{ka}/{2}}.
 \end{align}
 Lastly, we assume that each local bath is Ohmic, namely the density of states of the local bath can be written as
\begin{align}
D_{\rm bath}(\epsilon) \equiv 2\pi\sum_{\alpha}\delta(\epsilon-\omega_\alpha) = D_0 \epsilon \Theta(\epsilon).
\end{align}
Under this assumption and using (\ref{Apmdef}), the spectral function of the collection of baths is given by
\begin{align}
A_{\pm} (k,\mu) &= D_0 \left (\mu \mp  2t\cos \frac {ka}{2}\right ) \Theta\left (\mu \mp   2t\cos \frac {ka}{2}\right ).
\label{Assh}
\end{align}
Now we can determine the real space behaviour of the non-hermitian parameters by performing the Fourier transform
\begin{align}
\tilde \gamma_i (n) \equiv \frac{a}{2\pi}\int_{-\pi/a}^{\pi/a} {dk} e^{ikna}\gamma_i (k),
\label{tgammai}
\end{align}
where $n = 0,\pm 1,\pm 2,\cdots$ is the relative distance (in units of $a$) between two unit cells. It is clear from (\ref{gamma0}) and (\ref{gammaz}) that
\begin{align}
\tilde \gamma_0(n) = \frac{\kappa_1^2+\kappa_2^2}{\kappa_1^2-\kappa_2^2}\tilde \gamma_z(n).
\end{align}
Using (\ref{Assh}) and (\ref{gammax})-(\ref{gammaz}) in (\ref{tgammai}), we find
\begin{widetext}
\begin{align}
\label{gammaxx2}
\tilde\gamma_x(n) &= \frac{\kappa_1\kappa_2}{8\pi}D_0 t \left [4(\pi-\theta)\delta_{n,0} +2\pi (\delta_{n,1} + \delta_{n,-1})  + \frac{2\sin 2n\theta }{(4n^2-1)n}(1-\delta_{n,0})+\frac{\sin 2(n+1)\theta}{2n^2+3n+1} - \frac{\sin 2(n-1)\theta}{2n^2-3n+1} \right ] \\
\label{gammayx2}
\tilde\gamma_y(n) &= i\frac{\kappa_1\kappa_2 }{8\pi}D_0 t\left [2\pi (\delta_{n,1} - \delta_{n,-1})  + \frac{4\sin 2n\theta}{4n^2-1} -  \frac{\sin 2(n+1)\theta}{2n^2+3n+1}-\frac{\sin 2(n-1)\theta}{2n^2-3n+1} \right  ]\\
\label{gammazx2}
\tilde\gamma_z(n) &= - \frac{\kappa_1^2-\kappa_2^2}{8\pi}D_0 t \left  [4(\pi - \theta)\cos\theta \delta_{n,0} + \frac{2\sin 2n\theta \cos\theta}{(4n^2-1)n}(1-\delta_{n,0}) -\frac{4\cos 2n\theta \sin \theta}{4n^2-1} \right ],
\end{align}
\end{widetext}
where $\theta = 0$ for $\mu \geq 2t $ and  $\theta = \cos^{-1}(\mu/2t)$ for $\mu < 2t $. From (\ref{gammaxx2})-(\ref{gammazx2}) we immediately see that for small hopping amplitudes of the bath particles, more specifically for $t \leq \mu/2$, the dissipation of the system is restricted to on-site or to the nearest neighbour sites. Interestingly, the real-space non-Hermitian terms exhibit a type of universal behaviour in this parameter region, namely we see that $\gamma_i(n)/t$ are independent of the hopping parameter $t$ when $t \leq \mu/2$. For $t > \mu/2$, in contrast, the dissipation spreads beyond the nearest neighbour sites, which is expected due to the increasing mobility of the bath particles.

\section{Concluding remarks}
\label{sec:conclusion}
In this paper, we study one possible origin of the non-Hermitian Hamiltonians used in the non-Hermitian topological band theories, namely a quantum system coupled to an environment. We show that the effective non-Hermitian Hamiltonian of such a system will generally carry a constraint, which can affect the complex energy spectrum of the Hamiltonian in a significant way. In the Hermitian case, it is known that the topological band theory can be simulated by quantum systems such as cold atomic gases as well as by classical systems such as photonic crystals. For non-Hermitian topological band theories, however, our study suggests that their simulation using quantum systems may encounter some difficulties, due to the constraint on the complex energy spectrum of the resulting non-Hermitian Hamiltonian. On the other hand, it will be interesting to investigate the topological properties of such constrained non-Hermitian models. This will be the subject of future studies.

\acknowledgments
We thank Prof. Hui Zhai for initiating this project and for his valuable suggestions. J. M. Midtgaard wishes to acknowledge the support of the Danish Council of Independent Research -- Natural Sciences via Grant No. DFF - 4002-00336.

\end{document}